\documentclass[a4paper]{article}

\usepackage{INTERSPEECH2022}
\usepackage{xcolor}
\usepackage{subcaption}
\usepackage{graphicx}
\usepackage{tabu}
\usepackage{float}
\usepackage{enumitem}
\usepackage{makecell}
\usepackage{url}
\usepackage{multirow}
\newcommand*{\myDots}{\dots}

\makeatletter

\renewcommand\section{\@startsection {section}{1}{\z@}%
                                    {-2.1ex \@plus 0.0ex \@minus -.2ex}%
                                    {1.1ex \@plus.12ex}%
                                    {\normalfont\Large\bfseries\color{mybrown}}}
\renewcommand\subsection{\@startsection{subsection}{2}{\z@}%
                                      {-1.1ex \@plus 0.0ex \@minus -.2ex}%
                                      {0.4ex \@plus 0.12ex}%
                                      {\normalfont\large\bfseries\color{mybrownlighter}}}
\renewcommand\subsubsection{\@startsection{subsubsection}{3}{\z@}%
                                          {-0.75ex\@plus -0.1ex \@minus -.2ex}%
                                          {0.8ex \@plus .12ex}%
                                          {\normalfont\normalsize\bfseries\color{mybrown}}}

\newcommand{\thickhline}{%
    \noalign {\ifnum 0=`}\fi \hrule height 1pt
    \futurelet \reserved@a \@xhline
}
\newcolumntype{"}{@{\hskip\tabcolsep\vrule width 1pt\hskip\tabcolsep}}
\makeatother

\title{Multi-scale Speaker Diarization with Dynamic Scale Weighting}
\name{Tae Jin Park, Nithin Rao Koluguri, Jagadeesh Balam and Boris Ginsburg}
\address{NVIDIA}

\email{\{taejinp, nkoluguri, jbalam, bginsburg\}@nvidia.com}

\begin{document}

\maketitle

\begin{abstract}
Speaker diarization systems are challenged by a trade-off between the temporal resolution and the fidelity of the speaker representation. By obtaining a superior temporal resolution with an enhanced accuracy, a multi-scale approach is a way to cope with such a trade-off. In this paper, we propose a more advanced multi-scale diarization system based on a multi-scale diarization decoder. There are two main contributions in this study that significantly improve the diarization performance. First, we use multi-scale clustering as an initialization to estimate the number of speakers and obtain the average speaker representation vector for each speaker and each scale. Next, we propose the use of 1-D convolutional neural networks that dynamically determine the importance of each scale at each time step. To handle a variable number of speakers and overlapping speech, the proposed system can estimate the number of existing speakers. Our proposed system achieves a state-of-art performance on the CALLHOME and AMI MixHeadset datasets, with 3.92\% and 1.05\% diarization error rates, respectively. 

%The source code and models in this work will be open sourced through NeMo \cite{nemo_git}.

\end{abstract}
\noindent\textbf{Index Terms}: speaker diarization, multi-scale

\section{Introduction}
\label{sec:Introduction}
Speaker diarization is a task of partitioning an input audio stream into speaker-homogeneous segments allowing audio segments to be associated with speaker labels. Speaker diarization requires decisions on relatively short segments ranging from a few tenths of a second to several seconds. In speaker embedding extraction, to obtain high-quality speaker representation vectors, the temporal resolution should be sacrificed by taking a long speech segment. Thus, speaker diarization systems always face a trade-off between two quantities, i.e., the temporal resolution and quality of the representation. In the early versions of speaker diarization, the speaker homogeneous variable-length Mel-frequency cepstral coefficients (MFCCs) segments were generated through the detection of speaker change points~\cite{chen1998speaker}. 

% \cmt{downside of uniform seg}
Owing to the increased popularity of i-vectors~\cite{senoussaoui2013study} and x-vectors~\cite{sell2018diarization} in the field of speaker diarization, a uniform segmentation approach has been widely used, in which a fixed speaker segment length is applied to extract the speaker representation vector per segment. Because fixed-length segments lead to stable speaker representations by controlling the length factor in speaker embedding extraction, speaker diarization systems based on a uniform segmentation method~\cite{senoussaoui2013study,kenny2010diarization} have shown a competitive performance. However, the uniform segmentation approach has innate limitations in terms of the temporal resolution because reducing the segment length leads to a decrease in accuracy~\cite{park2021multi}. In a uniform segmentation setting, the shortest temporal resolution is limited to the hop-length during the segmentation process. Moreover, without a post processing approach, because each segment is assigned to only one speaker, the clustering approaches lack an overlap-aware diarization capability. 

%\cmt{modern diar sys - Pros} 
Recent speaker diarization systems such as an end-to-end approach \cite{fujita2019end} or target-speaker voice activity detection (TS-VAD)~\cite{medennikov2020target} employ feature frame-level speaker labels. These systems are based on sequence models such as long short-term memory (LSTM)~\cite{medennikov2020target, horiguchi2020end} or a Transformer-style encoder-decoder~\cite{fujita2019end}. Frame-level sequence model based systems enable a far superior temporal resolution of approximately 0.01 s. Another significant benefit of these methods is overlap-aware diarization where the neural network model outputs more than one speaker label output. The multi-dimensional sigmoid output enables an overlap detection without employing an additional resegmentation module on top of the diarization system. 

% \cmt{modern diar sys - Cons}
However, sequence model based methods also have a few drawbacks. First, because these sequence models are trained with a fixed number of speakers, sequence model based diarization systems often lack the capability to handle a flexible number of speakers \cite{medennikov2020target} or estimate their numbers~\cite{fujita2019end}. Second, according to the recent results of a diarization challenge \cite{watanabe2020chime, ryant2020third}, the performances of sequence model based end-to-end approaches~\cite{fujita2019end, horiguchi2020end, kinoshita2021integrating} still lag behind that of a state-of-art modular diarization~\cite{medennikov2020target, landini2022bayesian}. The hidden Markov model (HMM) based clustering approach proposed in \cite{landini2022bayesian} does not have the aforementioned downsides, having the capabilities of speaker counting and overlap detection and achieving a state-of-art performance. Thus, we compared our proposed system with the system proposed in \cite{landini2022bayesian} on the same datasets.

\begin{figure}[t]
\centerline{\includegraphics[width=0.98\linewidth]{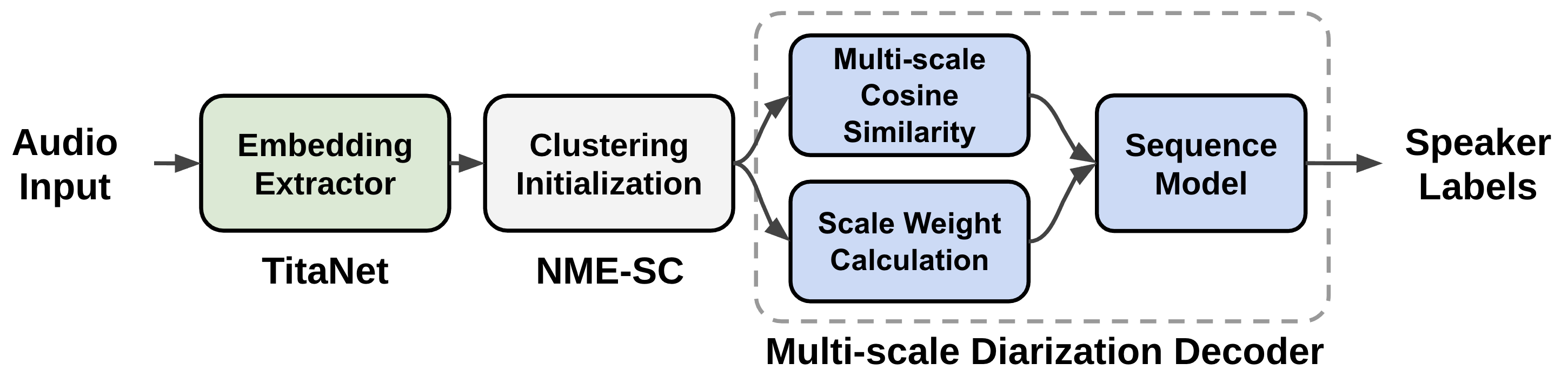}}
\vspace{-3.0ex}
\caption{Data-flow of the proposed multi-scale speaker diarization system.}
\label{fig:data_flow}
\vspace{-5.0ex}
\end{figure}
%%%
% \cmt{the proposed method} 
Our proposed multi-scale diarization decoder (MSDD) tackles the problems inherent to previous studies. The proposed MSDD was designed to support the following features: overlap-aware diarization, an improved temporal resolution, and a flexible number of speakers. As shown in Fig.\ref{fig:data_flow}, the proposed system comprises three components overall, i.e., a pretrained speaker embedding extractor (TitaNet, \cite{koluguri2021titanet}), a multi-scale speaker clustering module, and the MSDD model. 

Although we apply a multi-scale clustering idea from a previous study~\cite{park2021multi}, we do not estimate the static session level multi-scale weights. Instead, we propose the MSDD approach, which takes advantage of the initialized clusters by comparing the extracted speaker embeddings with the cluster-average speaker representation vectors. The weight of each scale at each time step is determined through a scale weighting mechanism where the scale weights are calculated from a 1-D convolutional neural network (CNN) applied to the input speaker embeddings and the cluster-average embedding based on the clustering results. Finally, based on the weighted cosine similarity vectors, an LSTM-based sequence model estimates thelabels of each speaker. As described in \cite{medennikov2020target}, using the clustering result as an initialization allows the proposed system to be free from permutations issue~\cite{fujita2019end} and turns the diarization problem into binary classification problem at each time step. Unlike the systems described in \cite{fujita2019end, medennikov2020target} we only train and test two-speaker models to handle a flexible number of speakers. Therefore, the speaker labels are predicted by taking the average of the sigmoid outputs from the multiple pairs. The 2-D dimensional output enables an overlap-aware diarization for the given input stream. 

In the experimental result section, we demonstrate that the proposed multi-scale approach is more accurate than our previous single scale (SS), clustering based diarization system. Furthermore, we show that our proposed system achieves new state-of-art results on the CALLHOME and AMI \textit{MixHeadset} datasets.
% In addition, the diarization performance is shown by comparing the proposed system with previously published results on CALLHOME and AMI dataset.  

\begin{figure}[t]
\centerline{\includegraphics[width=0.45\textwidth]{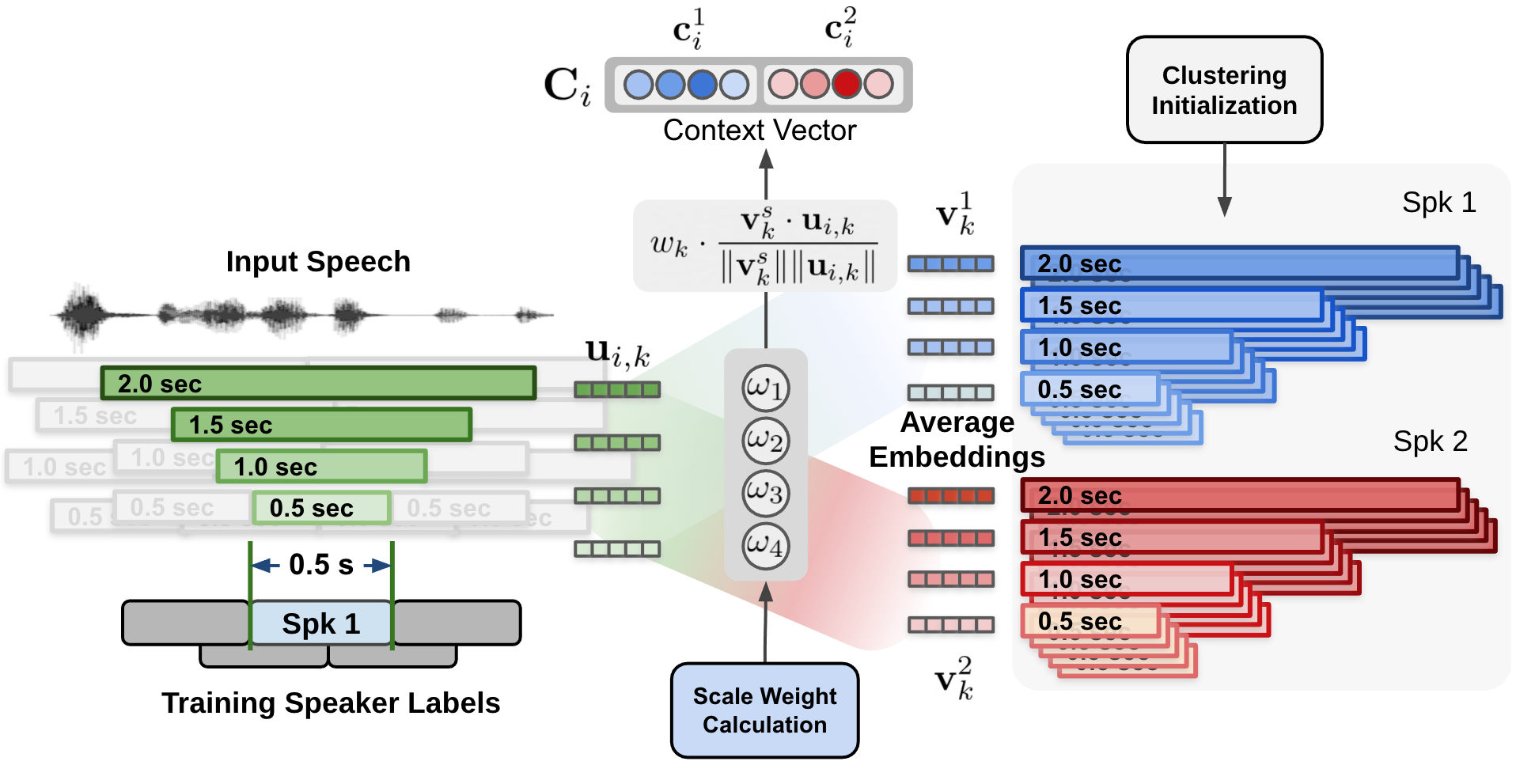}}
\vspace{-2.0ex}
\caption{Cosine similarity values from each speaker and each scale are weighted by the scale weights to form a weighted cosine similarity vector.}
\label{fig:weighted_sum}
\vspace{-5.0ex}
\end{figure}

\section{Multi-scale Clustering}
\label{subsec:multi-scale_clustering_system}

\subsection{Multi-scale segmentation}
\label{subsec:multi-scale segmentation}
The largest difference between a conventional single scale clustering method and multi-scale method is in the way in which the speech stream is segmented. The proposed system uses the same multi-scale embedding extraction and clustering mechanism as  in \cite{park2021multi}, except that the scale weights are computed differently in this work. We employ a uniform segmentation scheme that originally appeared in  \cite{senoussaoui2013study, kenny2010diarization}, and such segmentation is applied for multiple scales. Fig.~\ref{fig:weighted_sum} shows an example of a multi-scale segmentation. As shown in Fig.~\ref{fig:weighted_sum}, we refer to the finest scale, i.e., 0.5 s, as the base scale because the speaker label estimation is applied at the finest scale. We denote the number of scales as $K$.
% \cmt{mapping of multiple scales}

After segmentation is applied for all scales, grouping among the segments at each scale is then applied. The grouping process is conducted by assigning the segments from each scale for the corresponding base scale segment. The segments from the lower temporal resolution scales (in this example, 2.0, 1.5, and 1.0 s in length) are selected and grouped by measuring the distance between the centers of the segments and choosing the closest ones. By grouping the segments as in Fig. \ref{fig:weighted_sum}, each group will generate one weighted cosine similarity value when a weighted affinity matrix is calculated. A detailed description of multi-scale clustering can be found in \cite{park2021multi}. We use the speaker embedding extractor model proposed in \cite{koluguri2021titanet}, which generates 192-dimensional embedding.

\subsection{Clustering for initialization}
Because we previously proposed a multi-scale system in \cite{park2021multi}, we employ the auto-tuning spectral clustering approach proposed in \cite{park2019auto}, which is referred to as normalized maximum eigengap spectral clustering (NME-SC). From the clustering result, we obtain the estimated cluster label per each base scale segment and the estimated number of speakers $S$.
\begin{figure}[t]
\centerline{\includegraphics[width=0.88\linewidth]{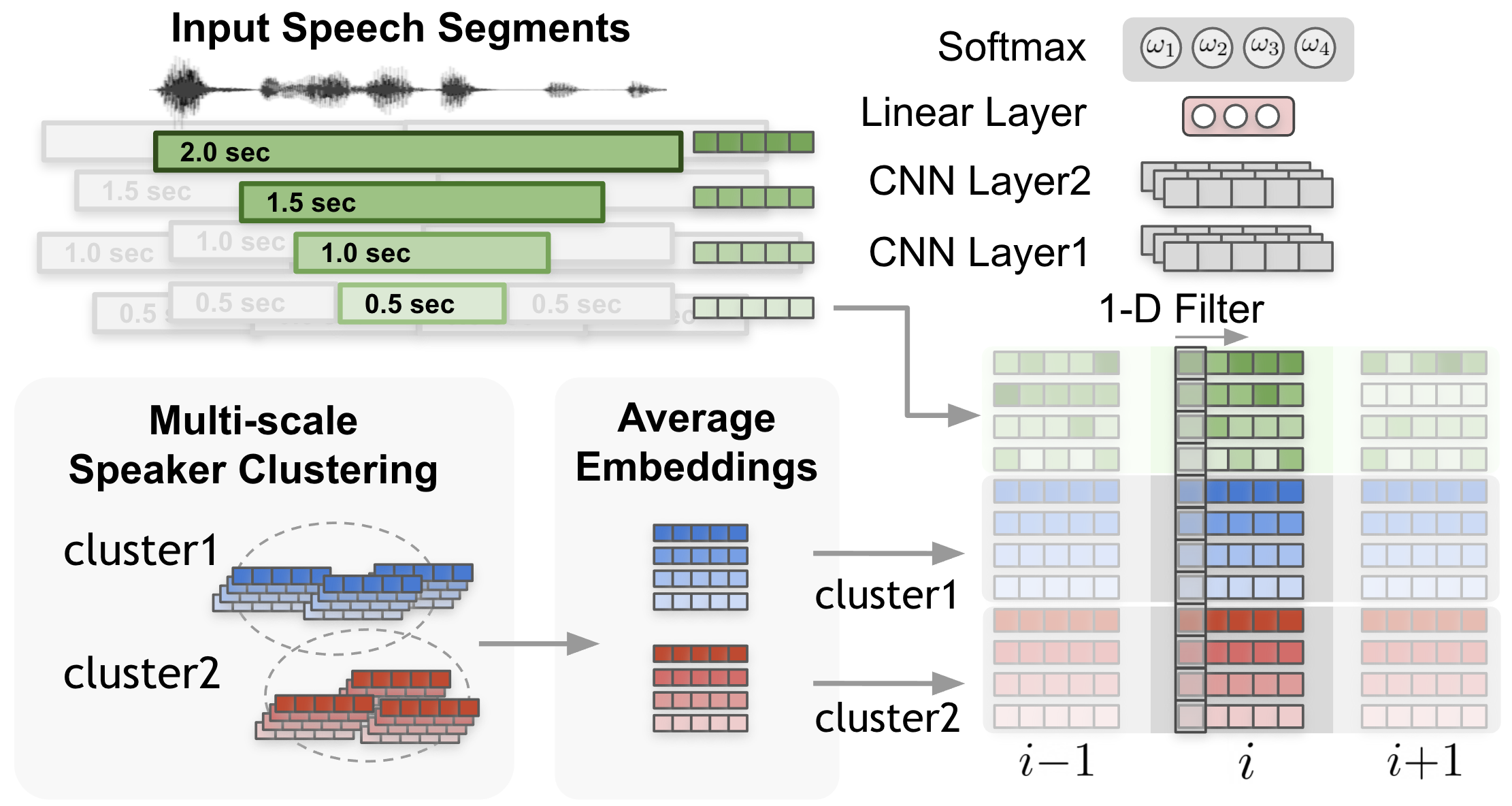}}
\vspace{-2.0ex}
\caption{Scale weights calculated from a 1-D CNN. The 1-D filter captures the context from the input embeddings and cluster-average embeddings. }
\vspace{-2.0ex}
\label{fig:scale_weighting_cnn}
\end{figure}

\begin{figure}[t!]
\centerline{\includegraphics[width=0.88\linewidth]{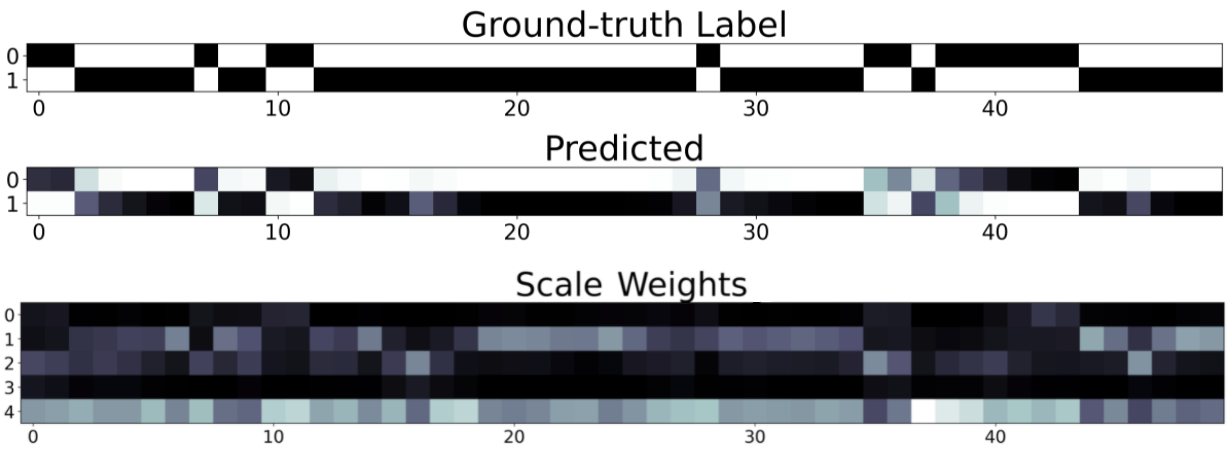}}
\vspace{-2.0ex}
\caption{Example plot of target labels, prediction, and scale weights (K=5). Note that scale weights vary at each time step.}
\label{fig:scale_weights}
\vspace{-4.0ex}
\end{figure}

% \subsection{Initialization of the cluster references}
\label{subsec:Initialization of the cluster references}
Based on the clustering results, we take the average of all the speaker embeddings obtained from the initial clustering result as in Eq.~(\ref{eq:mean_embedding}).  We obtain cluster-average embedding vectors $\textbf{v}_{k}^{s}$ from the initial clustering result as follows:
\begin{equation}
\label{eq:mean_embedding}
% \vspace{-1.0ex}
  \textbf{v}_{k}^{s} = \frac{1}{N_{k}}\sum_{i=1}^{N_k} \textbf{v}_{i,k}^{s},
\end{equation}
where $k$ is the scale index, $N_{k}$ is the number of $k$-th scale embeddings during the session and $s$ is the cluster index (speaker index) from the clustering result. 
\section{Multi-scale Diarization Decoder}

\subsection{1-D CNN for Dynamic Scale Weights}
\label{subsec:1-D CNN for Dynamic Scale Weights}
To dynamically adjust the scale weights during the inference phase, the proposed diarization system should look into the input speaker and reference embeddings simultaneously. Fig.~\ref{fig:scale_weighting_cnn} shows how scale weights are calculated in the case of $K$$=$$4$ scales for $S$$=$$2$ speakers. We stack the embeddings from the input signal and the average embeddings from the clustering results of two speakers ($s$$ = $$1$,$2$) in the following manner:
\begin{equation}
\label{eq:stacked_emb_input}
\vspace{-1.0ex}
\textbf{D}_{i} = [\textbf{u}_{i,0}; \text{\myDots}; \textbf{u}_{i,K-1};
                  \textbf{v}_{0}^{1}; \text{\myDots}\textbf{v}_{\text{$K$-$1$}}^{1}; 
                  \textbf{v}_{0}^{2}; \text{\myDots}; \textbf{v}_{K-1}^{2} ]^{\intercal},
\end{equation}
where $\textbf{u}_{i,k}$  are column-wise input speaker embeddings, and $\textbf{v}_{k}^{s}$ are cluster-average speaker embeddings, as described in Section~\ref{subsec:Scale Weighting Mechanism}. Thus, the CNN input becomes a $3K\times N_e$ matrix $\textbf{D}_{i}$, where $N_e$ is the embedding dimension, and $K$ is the number of scales.

To estimate the scale weight, we propose 1-D CNN with 1-D filters. The motivation behind using a 1-D filter is to compare the embedding vectors bin-by-bin such that the learned weights in the filters focus only on the difference between other embeddings for each bin. Two CNN layers are followed by two linear layers and the softmax layer, i.e., 
\begin{equation}
\label{eq:scale_weight_softmax}
    w_{k} = \frac{e^{f(z_{k})}}{\sum_{k=0}^{K-1} e^{f(z_{k})}},
\end{equation}
where $k$ is the scale index, and $f(z_{k})$ is the output of the linear layer (see Fig.~\ref{fig:scale_weighting_cnn}). The scale weight values $w_{k}$ in Eq.~ (\ref{eq:scale_weight_softmax}) are multiplied with the cosine similarity in an element-wise manner and create the context vectors for each speaker $\textbf{c}_{i}^s$. We concatenate these two speaker-specific context vectors for each $i$-th time step into the global context vector $\mathbf{C}_{i} = [\textbf{c}_{i}^{1}; \textbf{c}_{i}^{2}]$ of length $2K$
(see Fig.~\ref{fig:weighted_sum}).

\subsection{Scale Weighting Mechanism}
\label{subsec:Scale Weighting Mechanism}
% \cmt{the downside of clustering multi-scale diarization}
Although clustering based diarization shows a competitive performance, it lacks the ability to assign multiple speakers at the same time step, which is needed for overlap-aware diarization. In addition, the clustering-based multi-scale diarization requires a separate training step for multi-scale weights or a high-dimensional grid search. 
%These processes are time consuming and require a dataset to train or conduct a grid-search. 
This negatively affects the accuracy of the model for unseen domains during training. Moreover, the performance of a clustering-based multi-scale diarization system is heavily dependent on the scale weights, as shown in \cite{park2021multi}.

Hence, we derived a scale weighting mechanism that dynamically calculates the scale weights at every time step when a trained sequence model estimates the speaker labels. In the scale weighting system, during the training process, the scale weights are multiplied to the cosine similarity values to create a context vector that is fed to the sequence model (LSTM). The scale weighting mechanism is described in Fig. \ref{fig:weighted_sum}. Let $\textbf{u}_{i,k}$ be the $k$-th scale embedding of the input speech signal and $i$ be the time-step index. In addition, $\textbf{v}_{k}^{s}$ is the $k$-th scale average embedding vector of cluster $s$ from the initial clustering result in Eq.~(\ref{eq:mean_embedding}), and let ${w}_{k,i}$ be the $k$-th scale weight at the $i$-th time-step index. The scale weight $w_{k,i}$ is applied to the cosine distance between the inputs $\textbf{v}_{k}^{s}$ and $\textbf{u}_{i,k}$, i.e.,
\begin{equation}
  \textbf{c}_{i}^{s}[k] = w_{k} \cdot  \frac{\textbf{v}_{k}^{s}\cdot \textbf{u}_{i,k}  }{ \| \textbf{v}_{k}^{s} \| \| \textbf{u}_{i,k} \|},
\label{eq:scale_weight_multiplication}
\end{equation}
where $\textbf{c}_{i}^{s}$ is the context vector  for the $s$-th speaker (see Fig.\ref{fig:sequence_model}).

\subsection{Sequence model for speaker label estimation}
\label{subsec:Sequence Model for Speaker Label Estimation}
\begin{figure}[t]
\centerline{\includegraphics[width=0.68\linewidth]{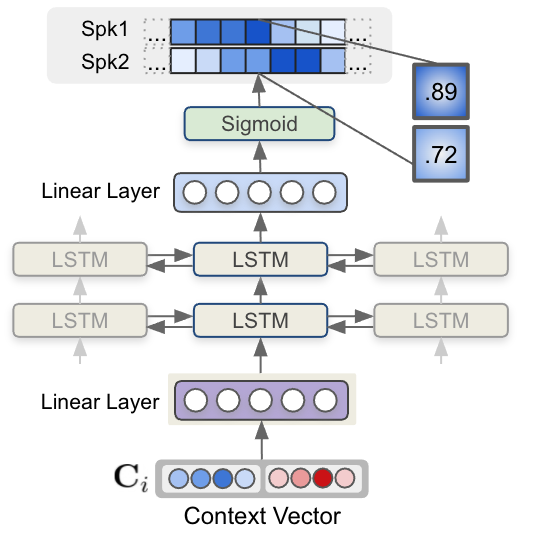}}
\vspace{-2.0ex}
\caption{Sequence modeling using LSTM for generating speaker labels.}
\vspace{-5.0ex}
\label{fig:sequence_model}
\end{figure}

The weighted cosine similarity vectors are fed to the sequence model. In our proposed system, we employ a two-layer Bi-LSTM. Fig.~\ref{fig:sequence_model} shows how the context vector is fed to the LSTM for estimating the speaker-wise label. Note that the output layer goes through the sigmoid layer such that the output value ranges from zero to 1 independently from the neighboring values. We use the binary cross-entropy loss to train the model. The binary ground truth label is generated by calculating the overlap between the ground truth speaker timestamps and the base scale segment: If the overlap is greater than 50\%, the segment is assigned a value of 1 (see Fig.\ref{fig:scale_weights}).

The speaker label inference for the sessions with more than two speakers is obtained by selecting pairs from the estimated number of speakers and averaging the prediction for the corresponding speaker from the two-speaker models.  For example, in the three speaker cases in which speakers A, B, and C exist, we average the two results for A from (A, B) and (A, C). Letting S be the total number of speakers in a session, then the proposed model infers $\binom{S}{2}$ number of speakers and then take average of all $S$$-$$1$ output pairs from a certain speaker through the following equation:
\begin{equation}
\label{eq:speaker_average}
    p(s,i) = \frac{1}{ \text{$S$ $-$ $1$}} \sum_{q \in U\text{$-$}\{s\} } \sigma(\mathbf{z})^{s, (s,q)}_{i}
\end{equation}
where $U$ is a complete set of the speakers, and $\sigma(\mathbf{z})^{s, (s,q)}_{i}$ is the sigmoid output of the $i$-th base scale segment of the $s$-th speaker, which is extracted from the pairing of the $s$-th and $q$-th speakers. We use a threshold value $T$ on the average sigmoid value $p(s,i)$ to obtain the final speaker label output for each base scale segment. If both sigmoid values are below the threshold $T$ for the given time step $i$, we then choose the label from the initial clustering result. 
% In addition, we only take up to two speakers with the largest sigmoid value at each time step $i$ because estimating the overlap of more than two speakers often becomes inaccurate. 

\section{Experimental Results}
\label{subsec:Experimental Results}

\subsection{Models and training}
We train the meeting and telephonic models (3.4M parameters) separately using the same dataset but different parameters. Table~\ref{tab:scale_info} shows the length of sclae and the speaker embedding extractor models. In this study,  we use TitaNet-M (13.4M parameters) and TitaNet-L (25.3M parameters) models \cite{koluguri2021titanet}. The shift lengths for each scale are half of each scale window length. The scale information in Table~\ref{tab:scale_info} is applied to both the neural network model and the clustering as initialization. The initial scale weights are set using the following equation:
\begin{equation}
\label{eq:initial_scale_weight}
     w_{k}  =  r  - \frac{(r-1)}{K-1}k,
\end{equation}
where the weights are linearly increased or decreased from the parameter value $r$ and the base scale (the highest index) is always assigned a value of $1.0$. In this way, the initial clustering scale weight vector $\mathbf{w}=[w_0, w_1, \dots, w_{K-1}]$ is formed. Note that the overall scale of the scale weights $w_k$ for clustering does not affect the result because the weighted sum of the cosine similarity values are min-max normalized during the clustering process \cite{park2019auto}. The proposed system requires parameter tuning on $r$ in Eq.~(\ref{eq:initial_scale_weight}) for clustering and threshold $T$ described in Section~\ref{subsec:Sequence Model for Speaker Label Estimation} for the MSDD.

\subsection{Evaluation setup}
An evaluation of a speaker diarization system depends on numerous conditions regarding the use of the development set, whether information on the oracle number of speakers is used, and the diarization error rate (DER) evaluation conditions such as collar and overlap inclusion. We only report the results based on the estimated number of speakers and oracle voice activity detection (VAD). We report two different DER evaluation settings following the name of the setups in \cite{landini2022bayesian}: (1) \textit{Forgiving, $\text{DER}_{A}$ }, 0.25-s collar and ignored overlap speech; and (2) \textit{Full, $\text{DER}_{B}$}, 0-s collar and included overlap speech.

\subsection{Datasets}

We use following datasets:
% \vspace{-1.0ex}
\begin{itemize}[leftmargin=2.5ex, noitemsep, topsep=0.2pt]
  \setlength{\parskip}{0pt}
  \setlength{\itemsep}{0pt plus 1.1pt}
% \subsubsection{CALLHOME (NIST-SRE-2000)} 
\item \textbf{CH(NS)}: NIST-SRE-2000 (LDC2001S97) is the most popular diarization evaluation dataset and referred to as CALLHOME in the speaker diarization papers. To make a fair comparison with other studies and our previous research, a two-fold cross validation is used for tuning the parameters using the split appeared in \cite{snyder2018x, snyder_git}.
\item \textbf{CH109}: Call Home American English Speech (CHAES, LDC97S42) is a corpus that contains only English telephonic speech data. We evaluate 109 sessions (CH109) which have two speakers per session. The parameters are optimized on the remaining 11 sessions in the CHAES dataset.
% \subsubsection{AMI meeting corpus} 
\item \textbf{AMI-MH}: AMI meeting corpus~\cite{mccowan2005ami} is a meeting speech dataset containing up to five speakers. For the evaluation, we use annotation files and the file lists of splits provided by \cite{bredin2020pyannote} from the \textit{MixHeadset} part. All parameters are optimized on the dev set. 

\item \textbf{The Fisher corpus}: The Fisher corpus contains 10 minute English conversations. We use Fisher corpus \cite{cieri2004fisher} to train the proposed model using the two-speaker setting. We use our own random split that has 10,000 sessions for training and 1,500 sessions for validation.

\end{itemize}
In terms of the hyperparameters, we use 256 nodes of hidden layer units for both hidden layers in the scale weight generation and the LSTM. We use 16 filters for the 1-D CNN described in \ref{subsec:1-D CNN for Dynamic Scale Weights}. F1 score is used for stopping the training or select the model. The results reported herein can be reproduced using the NeMo\footnote{https://github.com/NVIDIA/NeMo} open-source toolkit \cite{nemo_git}.

\begin{table}[t!]
\caption{Scale length and quantity for each model.}
\vspace{-4.0ex}
\small
\label{tab:scale_info}
\begin{center}
\begin{tabular}{ r| c | c | c } 
\Xhline{5\arrayrulewidth}
\textbf{Model}  & \multirow{2}{*}{$K$} & \multirow{2}{*}{\textbf{Scale (Segment) Length}} & \textbf{Emb.} \\
\textbf{Domain} &                    &                               & \textbf{Model} \\
\Xhline{1\arrayrulewidth}
Telephonic & 5 & [1.5, 1.25, 1.0, 0.75, 0.5] & TitaNet-M \\ 
Meeting    & 6 & [3.0, 2.5, 2.0, 1.5, 1.0, 0.5] & TitaNet-L \\
\Xhline{5\arrayrulewidth}
\end{tabular}
\end{center}
\vspace{-6.0ex}
\end{table}

\subsection{Diarization performance}

Table \ref{tab:der_table} shows the performance of the previously published speaker diarization performance on the same dataset. In Table \ref{tab:der_table}, note that $\text{DER}_{A}$ does not reflect the overlap detection capability. SS-Clus is our previous approach \cite{koluguri2021titanet}, where SS clustering-based diarization systems with TitaNet models~\cite{koluguri2021titanet} were used. The Equal-$\mathbf{w}$-Clus system uses equal weights, which means $w_{k}$$=$$1$ in Eq.~(\ref{eq:initial_scale_weight}) for all elements in the scale weight vector $\mathbf{w}$. Accordingly, Equal-$\mathbf{w}$-MSDD is the result obtained from the MSDD based on the initializing clustering with an equal scale weight. By contrast, Opt-$\mathbf{w}$-Clus indicates a clustering-based diarization result with the optimized scale weight vector $\mathbf{w}$ on each development set, and Opt-$\mathbf{w}$-MSDD refers to the system based on the clustering result Opt-$\mathbf{w}$-Clus. Whereas Opt-$\mathbf{w}$-MSDD shows an overall better result, we want to emphasize the importance of the Equal-$\mathbf{w}$-MSDD result because Equal-$\mathbf{w}$-MSDD does not require parameter tuning for the initializing clustering.

\begin{table}[t]
\caption{DER results from previous studies and the proposed methods. The number of speakers $N_s$ is estimated within the range $1$$\leq$$N_s$$\leq$$8$.}
\vspace{-4.2ex}
\small
\label{tab:der_table}
\begin{center}

\begin{tabular}{ r | c | c | c |  c  c  } 
\Xhline{5\arrayrulewidth}
                                   &       & \multicolumn{2}{c|}{Telephonic} &  \multicolumn{2}{c}{Meeting} \\
\Xhline{0.5\arrayrulewidth}
 \multirow{2}{*}{\textbf{Systems}} & Eval. & \textbf{CH}    &\textbf{CH}     & \multicolumn{2}{c}{\textbf{AMI-MH}} \\
                                   & Setup & \textbf{(NS)}     & \textbf{109}    & dev & test \\
\Xhline{3\arrayrulewidth}
Park \textit{et al.} \cite{park2021multi} & $\text{DER}_{A}$ & 6.46 & 2.04* & - & 3.32 \\ 
\Xhline{0.5\arrayrulewidth}
Aronow- \textit{et al.} \cite{aronowitz2020new} & $\text{DER}_{A}$ & 5.1 & - & - & - \\
\Xhline{0.5\arrayrulewidth}
Dawala- \textit{et al.} \cite{dawalatabad2021ecapa}& $\text{DER}_{A}$ & - & - & 2.43 & 4.03 \\ 
\Xhline{0.5\arrayrulewidth}
\multirow{2}{*}{Landini \textit{et al.} \cite{landini2022bayesian}} & $\text{DER}_{A}$ & 4.42 & - & 2.14 & 2.17 \\ 
                                                     & $\text{DER}_{B}$ & 21.77& - & 22.98 & 22.85 \\ 
\Xhline{3\arrayrulewidth}                                        
SS-Clus \cite{koluguri2021titanet} & $\text{DER}_{A}$ & 5.38 & 1.42 &  -  & 1.89 \\ 
\Xhline{0.5\arrayrulewidth}
Equal-$\mathbf{w}$-MS-Clus & $\text{DER}_{A}$                &  4.57         &  1.19          & 1.34          & 1.06 \\
\Xhline{0.5\arrayrulewidth}
\multirow{2}{*}{Equal-$\mathbf{w}$-MSDD} & $\text{DER}_{A}$  &  4.25         & \textbf{0.69}  & 1.34          & \textbf{1.05} \\
                        & $\text{DER}_{B}$                   &  20.39        & 10.94          & 22.21         & 21.18 \\
\Xhline{0.5\arrayrulewidth}
                Opt-$\mathbf{w}$-MS-Clus & $\text{DER}_{A}$  &  4.18         & 0.70           &\textbf{1.30}  & 1.06 \\
\Xhline{0.5\arrayrulewidth}
\multirow{2}{*}{Opt-$\mathbf{w}$-MSDD}   & $\text{DER}_{A}$  &  \textbf{3.92} & 0.73          &\textbf{1.30}  &1.06 \\
                                         & $\text{DER}_{B}$  &  20.14         & 10.82         & 22.20         & 21.19 \\
\Xhline{5\arrayrulewidth}
\end{tabular}
\end{center}
\vspace{-2.4ex}
{\scriptsize * This is \textit{eval} set in the CHAES dataset.}
\vspace{-5.0ex}
\end{table}

\section{Discussions}
% \subsection{Advantages and Disadvantages}

 Although our proposed system shows a competitive performance, there is still room for improvement, which requires further investigation. First, although the trained model can adjust the scale weight on the fly, to obtain improved results, we need to have separate models for both the meeting and telephonic data. Nevertheless, the proposed MSDD model does not show a significantly improved performance on the AMI datasets. We believe this can be overcome by including datasets from many different domains. Second, the temporal resolution is still limited to the shift length (0.25 s) of the finest scale. We observed that reducing the base scale to below 0.5 s only results in a larger error, which can also be an area of improvement.

Despite the aforementioned downsides, we found the following benefit of the MSDD. 
First, the proposed MSDD approach achieves similar or better results compared to clustering-based diarization results. Second, compared to the clustering based multi-scale diarization, the proposed system can achieve a more stable performance for unknown domains by using an equal scale weight. As investigated in previous multi-scale studies \cite{park2021multi}, the performance of clustering-based diarization results is heavily dependent on multi-scale weights. However, in terms of the proposed diarization decoder, the averaged speaker embeddings achieve a relatively minor effect from the scale weights used for the clustering. In particular, the smaller dependency on the tuned parameters is a benefit when the diarization system is deployed under real-life scenarios.
Third, the proposed system does not rely on an iterative procedure, unlike the system in \cite{medennikov2020target} where the speaker representation is extracted again after the first prediction. Thus, the proposed system can be applied to streaming diarization systems where we first apply clustering for a relatively short amount of time and then predict the diarization result on the incoming buffer in an incremental manner.

\section{Conclusions}
In this paper, we proposed a multi-scale diarization decoder with a scale weighting mechanism. The proposed system has the following benefits: First, this is the first study applying a multi-scale weighting concept with sequence model (LSTM) based speaker label estimation. Thus, the multi-scale diarization system enables overlap-aware diarization, which cannot be achieved with traditional clustering-based diarization systems. Moreover, because the decoder is based on a clustering-based initialization, the diarization system can deal with a flexible number of speakers. Finally, we showed a superior diarization performance compared to the previous published results. There are two future areas of research regarding the proposed system. First, we plan to implement a streaming version of the proposed system by implementing diarization decoder based on window-wise clustering. Second, the end-to-end optimization from speaker embedding extractor to diarization decoder can be investigated.

% \newpage
\bibliographystyle{IEEEtran}
\bibliography{mybib, refs}

\end{document}